\documentclass[pra,aps,twocolumn,superscriptaddress,showpacs]{revtex4}
\def\QED{\leavevmode\unskip\penalty9999 \hbox{}\nobreak\hfill
     \quad\hbox{\leavevmode  \hbox to.77778em{%
               \hfil\vrule   \vbox to.675em%
               {\hrule width.6em\vfil\hrule}\vrule\hfil}}
     \par\vskip3pt}
\def\qed{\leavevmode\unskip\penalty9999 \hbox{}\nobreak\hfill
     \quad\hbox{\leavevmode  \hbox to.77778em{%
               \hfil\vrule   \vbox to.675em%
               {\hrule width.6em\vfil\hrule}\vrule\hfil}}
\par\vskip3pt}
\def\ibb #1{\leavevmode\hbox{\kern.3em\vrule
     height 1.5ex depth -.1ex width .4pt\kern-.3em\rm#1}}
\def\Cx {{\ibb C}}

\newcommand{\be}{\begin{equation}}
\newcommand{\ee}{\end{equation}}
\newcommand{\ba}{\begin{array}}
\newcommand{\ea}{\end{array}}
\newcommand{\bqa}{\begin{eqnarray}}
\newcommand{\eqa}{\end{eqnarray}}

\newcommand{\tr}{\mbox{Tr}}

\begin{document}

\author{Zhihao Ma}
%\email[Email:]{mazhihao@sjtu.edu.cn}
\affiliation{Department of Mathematics, Shanghai Jiaotong
University, Shanghai, 200240, P.R.China}

\author{Fu-Lin Zhang}
\affiliation{Theoretical Physics Division, Chern Institute of
Mathematics, Nankai University, Tianjin, 300071, P.R.China}

\author{Jing-Ling Chen}
\email[Email:]{chenjl@nankai.edu.cn}\affiliation{Theoretical Physics
Division, Chern Institute of Mathematics, Nankai University,
Tianjin, 300071, P.R.China}

\date{\today}

\begin{abstract}
% insert abstract here

In this paper, we study metrics of quantum states. These metrics are
natural generalization of trace metric and Bures metric. We will
prove that the metrics are joint convex and contractive under
quantum operation. Our results can find important application in
studying the geometry of quantum states and is useful to detect
entanglement.
\end{abstract}

\title{Metrics Of Quantum States}
\pacs{03.67.-a, 03.65.Ta} \maketitle
\section{INTRODUCTION}
Let $S(H)$ be the set of all quantum states on a Hilbert space $H$.
Suppose we have two quantum states $\rho, \sigma$. The trace metric
is defined as: $d_{t}(\rho, \sigma)=\frac{1}{2}\tr|\rho- \sigma|$.
While the Uhlmann-Jozsa fidelity of  $\rho$ and $\sigma$ is defined
by $F(\rho, \sigma)=\tr \sqrt{\rho^{\frac{1}{2}}\sigma
\rho^{\frac{1}{2}}}$. And the Bures metric is defined as
$d_{B}(\rho, \sigma)=\sqrt{2-2F(\rho, \sigma)}$. Uhlmann-Jozsa
fidelity plays an important role in quantum information processing,
see \cite{Fid}.

Let $\rho$ be a quantum state. A positive-operator valued
measurement (POVM) is defined as a set of non-negative, Hermitian
operators $E_{k}$ which are complete in the sense that
$\sum_{k}E_{k}=I$. While projection measurement(PVM) also requires
that $E_{k}$ are all projections, see \cite{Nie1}.

Suppose we have two quantum states $\rho, \sigma$. A family of
metrics can be defined as follows:

{\bf Definition 1 \cite{Nico}.} Let $\rho, \sigma$ be quantum
states. And suppose $p$ is a fixed positive integer,  then we can
define a metric as follows: \begin{eqnarray}d_{p}(\rho,
\sigma)=\sup(\sum\limits^{N}_{k=1} |(\tr \rho
P_{k})^{\frac{1}{p}}-(\tr \sigma
P_{k})^{\frac{1}{p}}|^{p})^{\frac{1}{p}}\end{eqnarray} where the
supremum is taken over all finite families $\{P_{k}:k=1,2,...N\}$ of
mutually orthogonal projections  such that $\sum^{N}_{k=1}P_{k}=I$.
And we call the projections $P_{k}$ attaining the supremum  in the
equation (1) as optimal projections.

In \cite{Nico}, it is shown that when $p=1$, the metric $d_{1}(\rho,
\sigma)=2d_{t}(\rho, \sigma)=\tr|\rho- \sigma|$, that is,
$d_{1}(\rho, \sigma)$ equals two times the trace metric;  and when
$p=2$, the metric $d_{2}(\rho, \sigma)=d_{B}(\rho, \sigma)$, that
is, $d_{2}(\rho, \sigma)$ equals Bures metric.

The importance of the result of \cite{Nico} is that it unified two
important distances: trace metric and Bures metric in a common
frame.

In this paper, we will study the metrics introduced in \cite{Nico}
and get a family of " brother " metrics, proved a majorization
relation for metrics of states, also, our result lead to a new
fidelity measure and can be used to quantify entanglement.

\section{The Metrics of quantum states}

First, we want to prove that whether the metric $d_{p}(\rho,
\sigma)$ is contractive under quantum operation. Recall that a
quantum operation is a completely positive trace preserving (CPT)
map.

{\bf Theorem 1. (contractive of the  metric under CPT map)}  The
metric $d_{p}(\rho, \sigma)$ is contractive under quantum operation
for all $p$. That is, suppose $T$ is a completely positive trace
preserving (CPT) map, and $\rho, \sigma$ are density operators, then
we have the following inequality:
\begin{eqnarray}d_{p}(T(\rho), T(\sigma))\leq d_{p}(\rho, \sigma).\end{eqnarray}

{\bf Proof.} Note that $M_{n}$ becomes a Hilbert space with inner
product $<X,Y>:=\tr(X Y^{*}), X,Y\in M_{n}$. A linear map $T$
induces its adjoint map as: $<T(X),Y>=<X,T^{*}Y>$. If $T$ is a
positive map, then the adjoint map $T^{*}$ is also a positive map.
The trace preserving property of $T$ means that $T^{*}$ is unital,
that is, $T^{*}(I)=I$.

Now suppose $X_{k}$ are optimal projections for quantum states
$T(\rho), T(\sigma)$, so we get
$d_{p}(T(\rho),T(\sigma))=(\sum\limits_{k} |(\tr
T(\rho)X_{k})^{\frac{1}{p}}-(\tr
T(\sigma)X_{k})^{\frac{1}{p}}|^{p})^{\frac{1}{p}}$.

Let $Y_{k}=T^{*}(X_{k})$. Then $Y_{k}\geq 0$, and
$\sum\limits_{k}Y_{k}=\sum\limits_{k}T^{*}(X_{k})=T^{*}(\sum\limits_{k}(X_{k}))=T^{*}(I)=I$.

So we have  \begin{eqnarray*}&&\sum\limits_{k} |(\tr
T(\rho)X_{k})^{\frac{1}{p}}-(\tr
T(\sigma)X_{k})^{\frac{1}{p}}|^{p}\\
&=& \sum\limits_{k} |(\tr \rho T^{*}(X_{k}))^{\frac{1}{p}}-(\tr
\sigma T^{*}(X_{k}))^{\frac{1}{p}}|^{p}\\
&=& \sum\limits_{k} |(\tr \rho Y_{k})^{\frac{1}{p}}-(\tr
\sigma Y_{k})^{\frac{1}{p}}|^{p}\\
&\leq & d_{p}(\rho, \sigma)^{p} \end{eqnarray*}

So we get $d_{p}(T(\rho), T(\sigma))\leq d_{p}(\rho, \sigma)$.

{\bf Theorem 2.} $[d_{p}(\rho, \sigma)]^{p}$ is joint convex if and
only $p=1,2$, that is means, for $p\neq 1,2$, $[d_{p}(\rho,
\sigma)]^{p}$ is not joint convex.

The proof is left in appendix.

\section{a brother Metrics }

We know that for $p=1,2$, there are operational forms for
$d_{p}(\rho, \sigma)$, i.e., $d_{1}(\rho, \sigma)=\tr|\rho-
\sigma|$, and $d_{2}(\rho, \sigma)=\sqrt{2-2F(\rho, \sigma)}$, then
we can get the value of $d_{1}(\rho, \sigma)$ and $d_{2}(\rho,
\sigma)$ directly from the matrix entries of $\rho$ and $ \sigma$.
However, for other $p$, we can not enjoy this advantage, since
$d_{p}(\rho, \sigma)$ is defined via taking supremum of projections,
so a natural question arises: just like $p=1, 2$, can we get the
operational form for all $d_{p}(\rho, \sigma)$?

This problem is difficult, and we leave as a future topic. What we
want to say in this paper is that, we can introduce a new family of
metrics, these metrics can be seen as {\bf brother metrics} of the
metrics $d_{p}(\rho, \sigma)$, and have the advantages of easy to
calculate.

{\bf Definition 2.} Similar to definition 1, we define $D_{p}(\rho,
\sigma):=[\tr(|\rho^{\frac{1}{p}}-\sigma^{\frac{1}{p}}|^{p})]^{\frac{1}{p}}$.

{\bf Theorem 3.} Let $\rho, \sigma$ be quantum states, then
$D_{p}(\rho, \sigma)$ is a metric on $S(H)$.

{\bf Proof.} It is easy to show that  $D_{p}(\rho,
\sigma)=D_{p}(\sigma, \rho)$, $D_{p}(\rho, \sigma)\geq 0$ and
$D_{p}(\rho, \rho)= 0$. If $D_{p}(\rho, \sigma)= 0$, then
$|\rho^{\frac{1}{p}}-\sigma^{\frac{1}{p}}|=0$, so we get $\rho=
\sigma$. Recall that the Schattern p norm $||.||_{p}$ for an
operator $y$ is defined as (\cite{Pis}):
$||y||_{p}=[\tr(|y^{\frac{1}{p}}|^{p})]^{\frac{1}{p}}$. Now we
define two matrices $y_{1}$ and $y_{2}$, $y_{1}:=\rho^{\frac{1}{p}},
y_{2}:=\sigma^{\frac{1}{p}}$. Then use the triangle inequality for
the Schattern p norm $||.||_{p}$:
$$||y_{1}-y_{2}||_{p}\leq ||y_{1}||_{p}+||y_{2}||_{p}=2$$
$$||y_{1}-y_{3}||_{p}\leq ||y_{1}-y_{2}||_{p}+||y_{2}-y_{3}||_{p}$$
So we get the triangle inequality for $D_{p}$. Then $D_{p}(\rho,
\sigma)$ is a metric on $S(H)$, theorem is proved.

From $(UDU^{+})^{\frac{1}{p}}=UD^{\frac{1}{p}}U^{+}$, we get that
the metric $D_{p}(\rho, \sigma)$ is unitary invariant.

In quantum information theory, Majorization turned out to be a
powerful tool to detect entanglement. It was proved in Ref.
\cite{Nie2} that any separable state $\rho$ acting on
$\Cx^d\otimes\Cx^d$ is majorized by its reduced state $\rho_A$:
$$\rho_A \succ \rho  \quad\mbox{ i.e. }\quad \forall  k\leq d :
\sum_{i=1}^k \lambda_i^{(A)} \geq \sum_{i=1}^k \lambda_i,\nonumber
$$
where $\{\lambda_i\}$ and $\{\lambda_i^{(A)}\}$ are the decreasingly
ordered eigenvalues of $\rho$ respectively $\rho_A$.

Now We will give the following majorization relation for metric
$D_{p}(\rho, \sigma)$:

{\bf Theorem 4.} If $1\leq p\leq q$, and $\rho, \sigma$ are density
operators, define the vectors
$\lambda(|\sqrt[q]{\rho}-\sqrt[q]{\sigma}|^{q}),
\lambda(|\sqrt[p]{\rho}-\sqrt[p]{\sigma}|^{p})$ as the vectors of
eigenvalues for $|\sqrt[q]{\rho}-\sqrt[q]{\sigma}|^{q},
|\sqrt[p]{\rho}-\sqrt[p]{\sigma}|^{p}$, then the following
Majorization relation holds:
\begin{equation}\lambda(|\sqrt[q]{\rho}-\sqrt[q]{\sigma}|^{q})\prec_{w}
\lambda(|\sqrt[p]{\rho}-\sqrt[p]{\sigma}|^{p})\end{equation}

That is, $\lambda(|\sqrt[q]{\rho}-\sqrt[q]{\sigma}|^{q})$ is
majorized by $\lambda(|\sqrt[p]{\rho}-\sqrt[p]{\sigma}|^{p})$.

In particular, the following holds: \begin{equation}(D_{q}(\rho,
\sigma))^{q}\leq (D_{p}(\rho, \sigma))^{p}\end{equation}

{\bf Proof.}In \cite{Ando}, Ando proved that if $f(t)$ is a
non-negative, operator-monotone function on $[0,\infty)$ and
$|||.|||$ is a unitary invariant norm, then $$|||f(A)-f(B)|||\leq
|||f(|A-B|)|||, A,B\geq 0, $$  Or in majorization words
$$\lambda(|f(A)-f(B)|)\prec_{w} \lambda(f(|A-B|))$$

Since, for $1\leq p\leq q$, the function $f(t)=t^{\frac{p}{q}}$ is
operator-monotone, it follows from
\begin{equation}\lambda(|\rho^{\frac{p}{q}}-\sigma^{\frac{p}{q}}|)\prec_{w}
\lambda(|\rho-\sigma|^{\frac{p}{q}})\end{equation}

Consider the Schatten q-norm from equation (5) to get
\begin{equation}\tr(|\rho^{\frac{p}{q}}-\sigma^{\frac{p}{q}}|^{q})\leq \tr(|\rho-
\sigma|^{p})\end{equation}

And replace $\rho, \sigma$ in (6) by
$\rho^{\frac{1}{p}},\sigma^{\frac{1}{p}}$ respectively, then we
finished the proof.

Now we discuss the convex property of $D_{p}(\rho, \sigma)$. Similar
to the proof of theorem 2, we get the following:

$(D_{p}(\rho, \sigma))^{p}$ is joint convex if and only $p=1,2$.

Now we will discuss if the metrics $D_{p}(\rho, \sigma)$ is
contractive under quantum operation.

For $p=1$, we know that $D_{1}(\rho, \sigma)$ equal $d_{1}(\rho,
\sigma)$, so it is contractive under quantum operation.

For $p=2$, we know that $(D_{2}(\rho,
\sigma))^2=[\tr(|\rho^{\frac{1}{2}}-\sigma^{\frac{1}{2}}|^{2})]=2-2\tr(\rho^{\frac{1}{2}}\sigma^{\frac{1}{2}})$.

Uhlmann-Jozsa fidelity was widely studied, and plays a key role in
quantum information theory, but it is not easy to calculate, so some
alternative fidelity measures were introduced, see
\cite{Chen},\cite{Mis} and \cite{Mend}. The new fidelity introduced
in \cite{Chen},\cite{Mis} and \cite{Mend} are all proved to be a
good fidelity measure.

On the other hand, we know that Uhlmann-Jozsa fidelity can be
rewritten as $F(\rho, \sigma)=\tr
|\rho^{\frac{1}{2}}\sigma^{\frac{1}{2}}|$, that means, Uhlmann-Jozsa
fidelity is the trace of the modulus of the operator
$\rho^{\frac{1}{2}}\sigma^{\frac{1}{2}}$. however,
$\tr(\rho^{\frac{1}{2}}\sigma^{\frac{1}{2}})$ is exactly the trace
of the operator $\rho^{\frac{1}{2}}\sigma^{\frac{1}{2}}$. They only
differ from a phase factor!

So this leading to the following idea: if we define another
fidelity, called A-fidelity in this paper, as
\begin{eqnarray} F_A(\rho,\sigma)= [\tr(\sqrt{\rho}
\sqrt{\sigma})]^2.
\end{eqnarray}

Then we ask. can $F_A(\rho,\sigma)$ be a good fidelity measure?

The answer is yes. In fact, in \cite{Rag}, the author show that
$F_A(\rho,\sigma)$ has the following appealing properties:

{\bf Property1: CPT expansive property} if $\rho$ and $\sigma$ are
density matrices, $\Phi$ is a CPT map, then
$F_{A}(\Phi(\rho),\Phi(\sigma))\geq F_{A}(\rho,\sigma)$.

{\bf Property 2:} When $\rho=| \phi \rangle\langle \phi |$ and
$\sigma= | \varphi \rangle\langle \varphi |$ are two pure states,
Uhlmann-Jozsa fidelity and A-fidelity  both reduce to the inner
product, that is, $F(\rho,\sigma)=F_{A}(\rho,\sigma)=| \langle \phi
| \varphi \rangle |^2$.

Now we know that if $p=1$ or $p=2$, $D_{p}(\rho, \sigma)$ is joint
convex and also is contractive under quantum operation. We will show
that these are the {\bf only two cases} that satisfying CPT
contractive property.

For others $p\neq 1,2$, using the numerical method, we can get that
$D_{p}(\rho, \sigma)$ is neither decreasing nor increasing under
quantum operation.

We conclude as following:  $D_{p}(\rho, \sigma)$ is contractive
under quantum operation if and only $p=1, 2$.

To prove that for others $p$, $[d_{p}(\rho, \sigma)]^{p}$ is not
joint convex, we need the following simple example.

{\bf Example 1.} When $p\neq 1,2$, and Let $$\rho=\left (\begin{array}{cc} 0.2 & 0\\
0&0.8 \end{array}\right ), \;\;\;\;\;
\sigma=\left (\begin{array}{cc} 0.4 & 0\\
0&0.6\end{array}\right ), \;\;\;\;\;$$ We can prove that
$[d_{p}(\rho, \sigma)]^{p}$ is not joint convex. This example also
shows that $[D_{p}(\rho, \sigma)]^{p}$ is not joint convex for
$p\neq 1,2$.

\section{conclusions and  applications}

Our conclusion are the following:

1. For all $p$, $d_{p}(\rho, \sigma)$ is contractive under quantum
operation.

2. $[d_{p}(\rho, \sigma)]^{p}$ is joint convex if and only $p=1,2$.

3. $D_{p}(\rho, \sigma)$ is contractive under quantum operation if
and only $p=1, 2$.

4. $[D_{p}(\rho, \sigma)]^{p}$ is joint convex if and only $p=1,2$.

Since the metrics $d_{p}$ and $D_{p}$ are natural generalization of
trace metric and Bures metric, we wish that they can be used to
study the geometrical structure of quantum states, and find their
application in quantum information theory.

We know  that there are many entanglement measures, one of them is
the geometrical entanglement measure. Its idea is based on the
following: The set of all separable states is a convex set, denoted
as $S$, if we have a state $\rho$, then the closer the state  $\rho$
to the set $S$, the less entangled it is. So the entanglement
measure is defined as the minimal distance of the state $\rho$ to
any state of $S$:

$$E(\rho)=\min\limits_{\sigma\in S}D(\rho,\sigma).$$

Usually, we use Bures metric, that is, the metric $d_{2}$, to get
the geometrical entanglement measure, we wish that the metric
$D_{2}$ is also a good candidate for geometrical entanglement
measure, this work will be done in the future.

{\bf ACKNOWLEDGMENTS}

The authors thanks Prof. T.Ando for valuable discussions and great
help. The proof of theorem 1 and theorem 4 was with his help. The
authors also thanks for valuable discussions with Prof.E.Werner and
Prof.Q.Xu. This work is supported by the New teacher Foundation of
Ministry of Education of P.R.China (Grant No. 20070248087). J.L.Chen
is supported in part by NSF of China (Grant No. 10605013), and
Program for New Century Excellent Talents in University, and the
Project-sponsored by SRF for ROCS, SEM.

\appendix\section{proof of theorem 2}

Now we give the proof of theorem 2.

{\bf Proof.} For simplicity, we only discuss $2*2$ density matrices
$\rho, \sigma$, there is no difficulty to prove the  $N*N$ case when
apply the same method.

First, we will give an inequality. Let $P_{1},P_{2},P_{3},P_{4}$ be
$2$ dimension discrete probability distributions, Denote as
$P_{i}=\{P_{i1}, P_{i2}\}$, where $P_{i1},P_{i2}\geq 0,
P_{i1}+P_{i2}=1, i=1,2,3, 4$. Then the following holds: for all
$0\leq \lambda \leq 1$ and $p=1,2$, $$\sum\limits^{2}_{j=1}
|[\lambda P_{1j}+(1-\lambda)P_{2j}]^{\frac{1}{p}}-[\lambda
P_{3j}+(1-\lambda)P_{4j}]^{\frac{1}{p}}|^{p}$$$$\leq
\lambda\sum\limits^{2}_{j=1} |(
P_{1j})^{\frac{1}{p}}-(P_{3j})^{\frac{1}{p}}|^{p}+(1-\lambda)\sum\limits^{2}_{j=1}
|( P_{2j})^{\frac{1}{p}}-(P_{4j})^{\frac{1}{p}}|^{p}\eqno(8)$$

For $p=1$, the above inequality always holds, in fact, suppose
$P_{1}=\{a_{1}, 1-a_{1}\}$,$P_{2}=\{a_{2}, 1-a_{2}\}$,
$P_{3}=\{b_{1}, 1-b_{1}\}$, $P_{4}=\{b_{2}, 1-b_{2}\}$, then it
follows from the absolute values inequality $
|\lambda(a_{1}-b_{1})+(1-\lambda)(a_{2}-b_{2})|\leq \lambda
|a_{1}-b_{1}|+ (1-\lambda)|a_{2}-b_{2}|$.

For $p=2$, problem reduces to prove that the function
$f(a,b):=|a^{\frac{1}{2}}-b^{\frac{1}{2}}|^{2}+|(1-a)^{\frac{1}{2}}-(
1-b)^{\frac{1}{2}}|^{2}$ is joint convex, here $0\leq a,b\leq 1$. We
can get the Hessian
matrix of $f(a,b)$ as: $H(f)=\left (\begin{array}{cc} f^{''}_{aa} & f^{''}_{ab}\\
f^{''}_{ba}& f^{''}_{bb} \end{array}\right )$, and
$f^{''}_{aa}=\frac{1}{2}a^{-\frac{3}{2}}b^{\frac{1}{2}}
+\frac{1}{2}(1-a)^{-\frac{3}{2}}(1-b)^{\frac{1}{2}}$,
$f^{''}_{ab}=f^{''}_{ba}=-\frac{1}{2}a^{-\frac{1}{2}}b^{-\frac{1}{2}}
-\frac{1}{2}(1-a)^{-\frac{1}{2}}(1-b)^{-\frac{1}{2}}$,
$f^{''}_{bb}=\frac{1}{2}b^{-\frac{3}{2}}a^{\frac{1}{2}}
+\frac{1}{2}(1-b)^{-\frac{3}{2}}(1-a)^{\frac{1}{2}}$.

We know that the function $f(a,b)$ is joint convex if and only if
the Hessian matrix $H(f)$ is non-negative definite. And $H(f)$ is
non-negative definite if and only if $f^{''}_{aa}\geq 0,
f^{''}_{bb}\geq 0, f^{''}_{aa}f^{''}_{bb}-f^{''}_{ab}f^{''}_{ba}\geq
0$, $f^{''}_{aa}\geq 0, f^{''}_{bb}\geq 0$ always hold, so we only
need to prove $f^{''}_{aa}f^{''}_{bb}-f^{''}_{ab}f^{''}_{ba}\geq 0$,
This is equivalent to the following:
$a^{-\frac{3}{2}}b^{\frac{1}{2}}(1-a)^{\frac{1}{2}}(1-b)^{-\frac{3}{2}}+
a^{\frac{1}{2}}b^{-\frac{3}{2}}(1-a)^{-\frac{3}{2}}(1-b)^{\frac{1}{2}}\geq
2a^{-\frac{1}{2}}b^{-\frac{1}{2}}(1-a)^{-\frac{1}{2}}(1-b)^{-\frac{1}{2}}$,
And we know this holds from Cauchy-Schwarz inequality.

Now we have proved inequality (8) for $p=1,2$. We will use
inequality (8) to prove that for $p=1,2$, $(d_{p}(\rho,
\sigma))^{p}$ is joint convex.

Suppose $X_{k}$ are projections and $\sum\limits_{k}X_{k}=I$. Then
$\tr (\lambda \rho_{1}+(1-\lambda)\rho_{2})X_{k}, \tr (\lambda
\sigma_{1}+(1-\lambda)\sigma_{2})X_{k}$ and $
\tr(\rho_{1}X_{k}),\tr(\rho_{2}X_{k}), \tr(\sigma_{1}X_{k}),
\tr(\sigma_{2}X_{k})$ are all discrete probability distributions.

Put the four probability distributions
$\tr(\rho_{1}X_{k}),\tr(\rho_{2}X_{k}), \tr(\sigma_{1}X_{k}),
\tr(\sigma_{2}X_{k})$ in inequality (8), we get $\sum\limits_{j}
|[\lambda
\tr(\rho_{1}X_{k})+(1-\lambda)\tr(\rho_{2}X_{k})]^{\frac{1}{p}}-[\lambda
\tr(\sigma_{1}X_{k})+(1-\lambda)\tr(\sigma_{2}X_{k})]^{\frac{1}{p}}|^{p}\leq
\lambda\sum\limits_{j} |(
\tr(\rho_{1}X_{k}))^{\frac{1}{p}}-(\tr(\sigma_{1}X_{k}))^{\frac{1}{p}}|^{p}+(1-\lambda)\sum\limits_{j}
|(
\tr(\rho_{2}X_{k}))^{\frac{1}{p}}-(\tr(\sigma_{2}X_{k}))^{\frac{1}{p}}|^{p}
$ Then take supremum in both sides of the above inequality. Since
the optimal projection for $[d_{p}(\lambda
\rho_{1}+(1-\lambda)\rho_{2}, \lambda
\sigma_{1}+(1-\lambda)\sigma_{2}))]^{p}$ may not be the optimal
projection for $[d_{p}(\rho_{1},\sigma_{1})]^{p}$ and
$[d_{p}(\rho_{2},\sigma_{2})]^{p}$, we can get the needed result.
Theorem is proved.

\end{document}